\documentstyle[aps,manuscript,epsfig]{revtex}
\begin{document}
\title{Medium Effects in Kaon and Antikaon  Production in Nuclear Collisions
at Subthreshold Beam Energies}

\author{
F.~Laue$^a$, 
C.~Sturm$^b$,
I.~B\"ottcher$^d$,
M.~D\c{e}bowski$^e$,
A.~F\"orster$^b$,
E.~Grosse$^f$,
P.~Koczo\'n$^a$,
B.~Kohlmeyer$^d$,
M.~Mang$^a$,
L.~Naumann$^f$,
H.~Oeschler$^b$,
F.~P\"uhlhofer$^d$,
E.~Schwab$^a$,
P.~Senger$^a$,
Y.~Shin$^c$,
J.~Speer$^d$,
H.~Str\"obele$^c$,
G.~Surowka$^{a,e}$,
F.~Uhlig$^b$
A.~Wagner$^{a,*}$,
W.~Walu\'s$^e$\\
(KaoS Collaboration)\\ 
$^a$ Gesellschaft f\"ur Schwerionenforschung, D-64220 Darmstadt, Germany\\
$^b$ Technische Universit\"at Darmstadt, D-64289 Darmstadt, Germany\\
$^c$ Johann Wolfgang Goethe Universit\"at, D-60325 Frankfurt am Main, Germany\\
$^d$ Phillips Universit\"at, D-35037  Marburg, Germany\\
$^e$ Jagiellonian University, PL-30-059 Krak\'ow, Poland\\
$^f$ Forschungszentrum Rossendorf, D-01314 Dresden  and
Technische Universit\"at Dresden, Germany\\
$^*$ Present address:
NSCL Michigan State University, East Lansing, MI 48824-1321, USA
}
\maketitle


\begin{abstract}
Production cross sections of 
K$^+$ and K$^-$ mesons have been measured in C+C collisions  at beam energies 
per nucleon below and near the nucleon-nucleon threshold. 
At a given beam energy, the  spectral slopes of the 
K$^-$ mesons are significantly steeper than the 
ones of the K$^+$ mesons. 
The excitation functions for  K$^+$ and K$^-$ mesons nearly coincide 
when correcting for the threshold energy. 
In contrast,  the K$^+$ yield exceeds the K$^-$ yield by a factor of about 100 
in proton-proton collisions at beam energies  near the respective 
nucleon-nucleon thresholds.
\end{abstract}

\newpage
The properties of strange mesons in a medium of finite baryon density are 
essential for our understanding of the strong interaction.
According to various theoretical approaches, antikaons  
feel strong attractive forces in the nuclear medium whereas 
the in-medium kaon-nucleon potential is expected to be slightly repulsive
\cite{kaplan,brown1,schaffner,waas,lutz}. Predictions 
have been made that the effective 
mass of the  K$^-$ meson decreases with increasing nuclear density
leading to K$^-$ condensation in neutron stars above 3
times saturation density $\rho_0$. This effect is 
expected to influence significantly  the evolution of supernova explosions: 
the K$^-$ condensate softens the nuclear equation of state  
and thus causes a core with 1.5 - 2 solar masses to collapse 
into a black hole rather than to form a neutron star \cite{brobet,li_lee_br}.

Experimental evidence for the attractive in-medium K$^-$N potential was found
in K$^-$ nucleus scattering \cite{friedman1} 
and kaonic atoms \cite{friedman2}.
Strong effects are expected in relativistic nucleus-nucleus collisions 
where baryonic densities of several times the saturation density 
$\rho_o$   can be reached. Under these conditions the 
K$^-$ effective mass will be reduced and thus 
the kinematical threshold for the process NN$\to$K$^-$+K$^+$+NN      
(which in free space corresponds to a kinetic beam energy of 2.5 GeV) 
will be lowered. As a consequence, the  
K$^-$ yield in A+A collisions at bombarding energies below the NN
threshold will be enhanced significantly as compared to the case 
without in-medium mass reduction. 
In contrast, the yield of K$^+$ mesons is predicted to be decreased
as the K$^+$ effective mass and thus the in-medium K$^+$ production
threshold is slightly increased \cite{schaffner,li_ko_fang,cassing}.
The in-medium KN potentials are also expected to affect the 
propagation of kaons and antikaons hence modifying 
their  emission pattern in nucleus-nucleus collisions.  

According to these  considerations, the K$^-$/K$^+$ ratio
observed in nucleus-nucleus collisions  at beam energies below the 
NN threshold 
is sensitive to the in-medium properties of kaons and antikaons.
Moreover, relativistic transport calculations find distinct differences in the 
K$^+$ and K$^-$ spectral slopes, again  due to medium effects. 
In particular, these 
calculations predict that the K$^-$ spectra fall off 
steeper than the K$^+$ spectra due to the decrease of the K$^-$ effective 
mass in the nuclear medium  \cite{bra_ca_mo}.   

Recent experimental studies of kaon and antikaon production in  Ni+Ni collisions
found a large  K$^-$ yield  at 1.8 AGeV 
\cite{schroeter,barth}, a vanishing in-plane flow of K$^+$ mesons 
\cite{ritman} and an enhanced out-of-plane emission of K$^+$ mesons 
\cite{shin}. These observations  are indications for    
in-medium modifications of the K meson properties 
\cite{li_ko_fang,cassing,li_flow,li_ko_br}.

In this Letter we report on the first comparative measurement of 
the K$^-$ and K$^+$ excitation functions in nucleus-nucleus collisions 
at subthreshold beam energies (i.e. at beam energies per nucleon 
below the NN threshold).   
We have chosen the system C+C at bombarding energies between 0.8 and 2 AGeV. 
For each energy, differential cross sections for 
K meson production were obtained  at various emission angles.  
In this light collision 
system, the loss of K$^-$ mesons via the strangeness exchange
reaction K$^-$N$\to$Y$\pi$ (with Y = $\Lambda,\Sigma$) 
and its effect on the K$^-$ spectrum should be significantly smaller 
than in the Ni+Ni system \cite{schroeter,barth}.  

The experiment was performed with the Kaon Spectrometer (KaoS) at the 
heavy ion synchrotron (SIS) at GSI in Darmstadt \cite{senger}. 
The  $^{12}$C beam had an intensity of about 1$\times10^8$ ions per spill. 
The energy loss of the projectiles  
in the target (thickness 5 mm) is less than 5 MeV per nucleon.
This magnetic spectrometer has a large acceptance
in solid angle and momentum ($\Omega\approx$30 msr, $p_{max}/p_{min}\approx$2).
The short distance of 5 - 6.5 m from target
to focal plane minimizes kaon decays in flight. 
Particle identification and the trigger are based on separate measurements of
velocity, momentum and time-of-flight. The trigger suppresses pions and
protons by factors of 10$^2$ and 10$^3$, respectively.
The background due to spurious tracks and pile-up is
removed by trajectory reconstruction based on
three large-area multi-wire chambers. The  remaining background below
the kaon mass peak ($\approx$ 20 \%) is subtracted. The loss of kaons
decaying in flight is determined (and corrected)
by Monte Carlo simulations using the GEANT code.
The K mesons were registered
at polar angles of $\Theta_{lab}$ = 32$^{\circ}$ - 70$^{\circ}$
over a momentum range of 260 $<$ p$_{lab}<$ 1200 MeV/c.
The (approximate) raw numbers of detected K mesons are listed in Table 1.

Figure 1 shows the invariant inclusive cross section for K$^+$ (open symbols) 
and K$^-$ production (full symbols)  
as function of the center-of-mass kinetic energy 
measured in C+C collisions at 0.8, 1.0, 1.2, 1.5, 1.8 and 2 AGeV under different
laboratory angles. 
The error bars shown are due to statistics only. 
An overall systematic error of 10 \%
due to efficiency corrections and normalization procedures has to be added. 
The lines represent 
Boltzmann distributions d$^3\sigma$/dp$^3 \propto$ exp(-E/T) fitted 
to the spectra (only some of the fits are shown). 
The resulting inverse slope parameters T 
are given in Table~1.
The T-values for the antikaons are significantly  
smaller than the ones for the kaons. 

From the K$^+$ spectra taken at 1.0 and 1.8 AGeV
one can deduce a polar angle distribution 
which turns out to be  forward-backward peaked. The nonisotropic
contribution to the kaon yield is about 20\%.
This value agrees with the prediction of relativistic transport calculations
for C+C collisions at 1.8 and 2.0 AGeV. 
Inclusive production cross sections $\sigma_K$ 
are calculated from the extrapolations in c.m. energy and angles.
The kaon and antikaon spectra are parameterized by Boltzmann distributions
in accordance with the data in fig.1, with the results of other 
measurements \cite{schroeter,barth,ahner}
and with the results of model calculations \cite{bra_ca_mo}. 
A nonisotropic contribution of 20 \% is taken into account both for 
K$^+$ and K$^-$ mesons  although the transport calculations
predict a value of 40 \% for the antikaons. We do not apply this 
correction to the data but rather include this uncertainty into 
the systematic error of the antikaon yields.

In Fig. 2 we present the K$^+$ and K$^-$ multiplicities per participating
nucleon for C+C collisions as a function of the  energy above threshold 
($\sqrt s$-$\sqrt s_{th}$) in the nucleon-nucleon (NN) system.
The values of $\sqrt s$-$\sqrt s_{th}$ 
are calculated according to 
$s$ = (E$_{pro}$+2m$_N$)$^2$ - p$_{pro}^2$ with E$_{pro}$ and p$_{pro}$ 
the projectile kinetic  energy and momentum per nucleon and m$_N$ 
the nucleon mass. The threshold energy
is $\sqrt s_{th}$=m$_K$+m$_{\Lambda}$+m$_N$ = 2.55 GeV for K$^+$ production and 
$\sqrt s_{th}$=2m$_K$+2m$_N$ = 2.86 GeV for K$^-$K$^+$ pair production. 
K meson multiplicities are defined as M$_K$ =$\sigma_K/\sigma_R$ 
with  $\sigma_R$ the geometrical cross section of the reaction which is 
$\sigma_R$ = 4$\pi$(1.2 A$^{1/3}$)$^2$ fm$^2$ = 0.94 b for C+C. 
The average number of nucleons which 
participate in a reaction 
is $<$A$_{part}>$ = A/2 = 6 for C+C (according to a 
geometrical model \cite{huefner}). For comparison, Fig. 2 presents also the 
parameterizations of the isospin averaged  cross sections for K meson production
in nucleon-nucleon collisions \cite{sibirtsev,brat_cass,si_ca_ko}. 
These calculations reproduce the available 
experimental cross sections  including the most recent data measured at 
COSY close to threshold \cite{balewski}. 
The multiplicities as shown in Fig. 2 are calculated  from 
the elementary cross sections using $\sigma_R$ = 45 mb 
and A$_{part}$=2 for nucleon-nucleon collisions.

The data in Fig. 2 demonstrate that the excitation functions 
for K$^+$ and K$^-$ production in C+C collisions are quite similar 
when correcting the energy axis  for the threshold energies. 
This similarity is also observed when the invariant differential cross sections
for K$^+$ and K$^-$ production are compared at equivalent beam energies
(see Fig.1). In nucleon-nucleon (NN)
collisions, however, the K$^+$ yield exceeds the K$^-$ yield by 1-2 orders of 
magnitude for beam energies close to threshold.
The comparison of the K$^+$ and K$^-$ excitation functions for C+C and
nucleon-nucleon reactions near threshold clearly indicates that
in the nuclear medium the K$^-$/K$^+$ ratio is much larger than in
nucleon-nucleon interactions.
 
In nucleus-nucleus collisions, strange mesons can also be produced in secondary
processes like $\pi$N$\to$K$^+$Y, $\Delta$N$\to$K$^+$YN, 
$\pi$N$\to$K$^-$K$^+$N,  $\Delta$N$\to$K$^-$K$^+$NN and $\pi$Y$\to$K$^-$N. 
According to relativistic transport models 
the pion and $\Delta$ induced
sequential processes dominate the K$^+$ production at bombarding energies
near the kinematical threshold \cite{cassing,fuchs}. 
In the case of K$^-$ mesons, however,  
the measured yield is still about 6 times larger than 
what is predicted to emerge from 
all pion induced processes (including the strangeness exchange
reaction $\pi$Y$\to$K$^-$N). In this comparison 
the in-medium mass modifications are omitted  
but absorption is taken into account \cite{cassing}.

Transport calculations predict different 
spectral slopes for K$^-$ and K$^+$ mesons
as a result of the differences in the effective masses 
\cite{bra_ca_mo,cass_brat}. 
Figure 3 presents the ratio of the 
invariant production cross sections (K$^-$/K$^+$)  
as a function of the c.m. kinetic energy
in C+C collisions at 1.8 AGeV. In such a representation
the systematic uncertainties of the experimental data  are partly compensated  
and the difference in slope is clearly visible.
The K$^-$/K$^+$ ratio steeply 
decreases with increasing kinetic energy of the K mesons. 
This is not a trivial observation 
as K$^-$ mesons with low energies are  more strongly affected by absorption than
K$^-$ mesons with  higher energies \cite{dover}.
Relativistic transport calculations predict a constant  K$^-$/K$^+$ ratio 
when  in-medium mass modifications 
of the K mesons are neglected \cite{bra_ca_mo}.
If the calculations  take into account in-medium effects, both the yields and 
the spectral shapes of the K mesons are modified. 
In this case the K$^-$ yield at low
momenta is enhanced by a factor of about 5 (e.g. at E$_{cm}^{kin}$= 0.1 GeV)
mainly due to the reduced K$^-$ effective mass which drops according to 
m$^*$ = m$^0$ (1 - 0.24 $\rho/\rho_0$) \cite{bra_ca_mo}.
This effect overcompensates the 
absorption of K$^-$ mesons by strangeness exchange reactions 
K$^-$N$\to$Y$\pi$ which  are expected to reduce the K$^-$ yield predominantly
at low kaon momenta \cite{dover}.

In Ni+Ni collisions at 1.8 AGeV, a  K$^+$/K$^-$ cross section ratio of
21$\pm$9 was found \cite{barth}.
According to Table~1 the corresponding ratio 
is 39$\pm$6  in C+C collisions.
This result is in contrast to the  expectation that 
in the light system the absorption of K$^-$ mesons should be reduced
(the K$^+$ meson cannot be absorbed in nuclear matter due
to its $\overline s$ quark content).       
The observation of a decreasing
K$^+$/K$^-$ ratio with increasing size of the collision system may be 
partly caused by phase space effects: at 1.8 AGeV, K$^+$ production is 
above threshold whereas K$^-$ production is below, and the probability 
for subthreshold particle production via multiple collisions  
rises with increasing number of participants. 
Another reason could be the density dependence of the K$^-$ effective mass.
An increasing system size corresponds to an increased average baryonic
density which in turn may cause a more strongly reduction of the K$^-$ effective 
mass and thus a lower effective K$^-$ production threshold. 
The resulting relative K$^-$ yield is enhanced in the heavier system 
because the absorption is overcompensated by the lowered threshold. 

In summary, we have measured  K$^+$ and K$^-$ production in
C+C collisions at beam energies below and near the kinematical threshold.   
The excitation functions nearly coincide when correcting for the
threshold energy.  
In contrast, the K$^-$/K$^+$ ratio measured in p+p collisions is about 0.01
for beam energies near threshold. The spectral slopes of the K$^-$ mesons 
are steeper than the ones of K$^+$ mesons at the same beam energy. 
Within the framework of transport 
calculations, both
the enhanced K$^-$ yield and the steep K$^-$ spectral slope is explained
by a reduced effective mass of the  K$^-$ meson in the nuclear medium.  

This work was supported by the German Federal Government (BMBF), by the
Polish Committee of Scientific Research (Contract No. 2P03B11515) and 
by the GSI fund for University collaborations.

\begin{table}
Table 1: Sample sizes, inverse slope parameters and inclusive production 
cross sections
for kaons and antikaons in C+C collisions. The values for T and $\sigma$ 
are determined 
by fitting a Boltzmann distribution d$^3\sigma$/dp$^3$ $\propto$ exp(-E/T)
to the data. In the extrapolation to full phase space 
the nonisotropic angular distribution is taken into account (see text).     
The quoted errors on the cross sections include systematic effects.
\begin{center}
\begin{tabular}{|c|c|c|c|c|}
beam energy & $\Theta_{lab}$ &N(K$^+$)&T(K$^+$)&$\sigma$(K$^+$)\\
(AGeV) &  & & (MeV) & (mb)\\
\hline
0.8  & 44$^{\circ}$   & 740 & 54$\pm$3 & 0.015$\pm$0.003 \\
1.0  & 44$^{\circ}$, 54$^{\circ}$ & 1200   & 58$\pm$6 & 0.1$\pm$0.02 \\
1.2  & 40$^{\circ}$  & 2400 & 64$\pm$4 & 0.3$\pm$0.05 \\
1.5  & 32$^{\circ}$, 48$^{\circ}$  &7000 & 74$\pm$5 & 1.3$\pm$0.2 \\ 
1.8  & 32$^{\circ}$, 40$^{\circ}$, 48$^{\circ}$, 60$^{\circ}$ &17000 & 75$\pm$5 & 3.0$\pm$0.3 \\
2.0  & 32$^{\circ}$, 40$^{\circ}$  &11000 & 78$\pm$5 & 5.0$\pm$0.5  \\
\hline
\hline
beam energy & $\Theta_{lab}$ &N(K$^-$)&T(K$^-$)& $\sigma$(K$^-$)\\
(AGeV) &  & & (MeV) & (mb)\\
\hline
1.5  & 40$^{\circ}$  &50    & 40$\pm$15 & 0.016$\pm$0.006 \\ 
1.8  & 40$^{\circ}$, 60$^{\circ}$ & 2000 & 55$\pm$6  &  0.076$\pm$0.02 \\
2.0  & 40$^{\circ}$  &370   & 55$\pm$7  & 0.19$\pm$0.06 \\
\end{tabular}
\end{center}
\end{table}

\newpage       
Figure 1:
Inclusive invariant  production cross-sections of K$^+$ mesons (open symbols) 
and K$^-$ mesons (full symbols)  measured in
C+C collisions at beam energies of 2, 1.8, 1.5, 1.2, 1.0 and 0.8 AGeV
(from top to bottom)  under different  laboratory angles
(as indicated).  
The solid lines  correspond to  Boltzmann distributions fitted to 
some of the spectra (see text and Table 1).

\vspace{1.cm}
Figure 2:

Kaon and antikaon multiplicity per participating nucleon
as a function of the Q-value  for C+C 
collisions (open squares: K$^+$, full dots: K$^-$) and 
nucleon-nucleon  collisions.
The error bars include systematic effects. 
The lines correspond to parameterizations of the  isospin averaged 
cross sections for K meson production measured in proton-proton collisions 
(full line: K$^+$, dashed line: K$^-$) 
\protect\cite{sibirtsev,brat_cass,si_ca_ko}.

\vspace{1.cm}
Figure 3:

K$^-$/K$^+$ ratio as a function of the center-of-mass kinetic energy  
measured in C+C collisions at 1.8 AGeV and at $\Theta_{lab}=40^0$.

\begin{figure}[h]
\hspace{1.0cm}\mbox{\epsfig{file=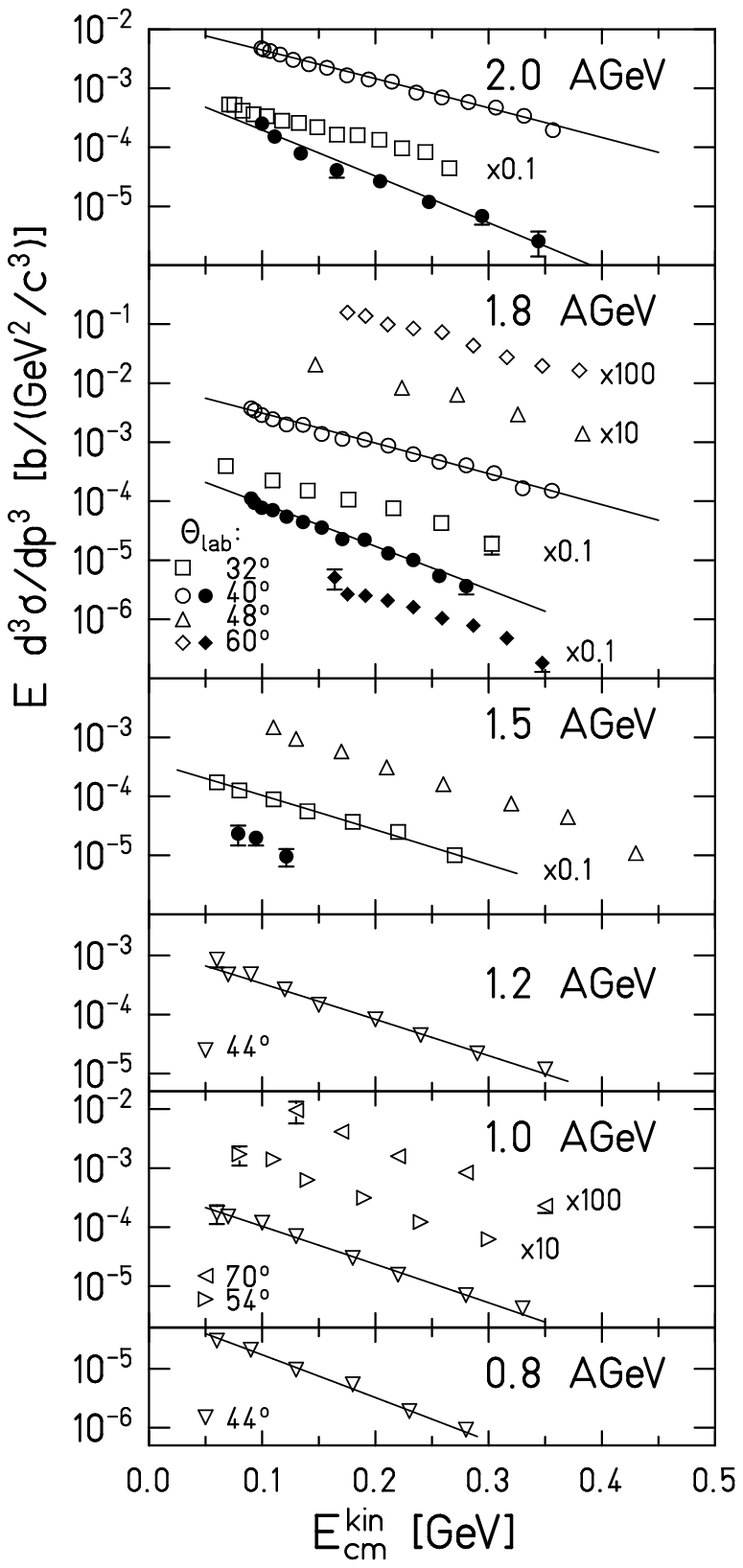,width=13.cm}}
\caption{}
\end{figure}

\begin{figure}[h]
\hspace{1.0cm}\mbox{\epsfig{file=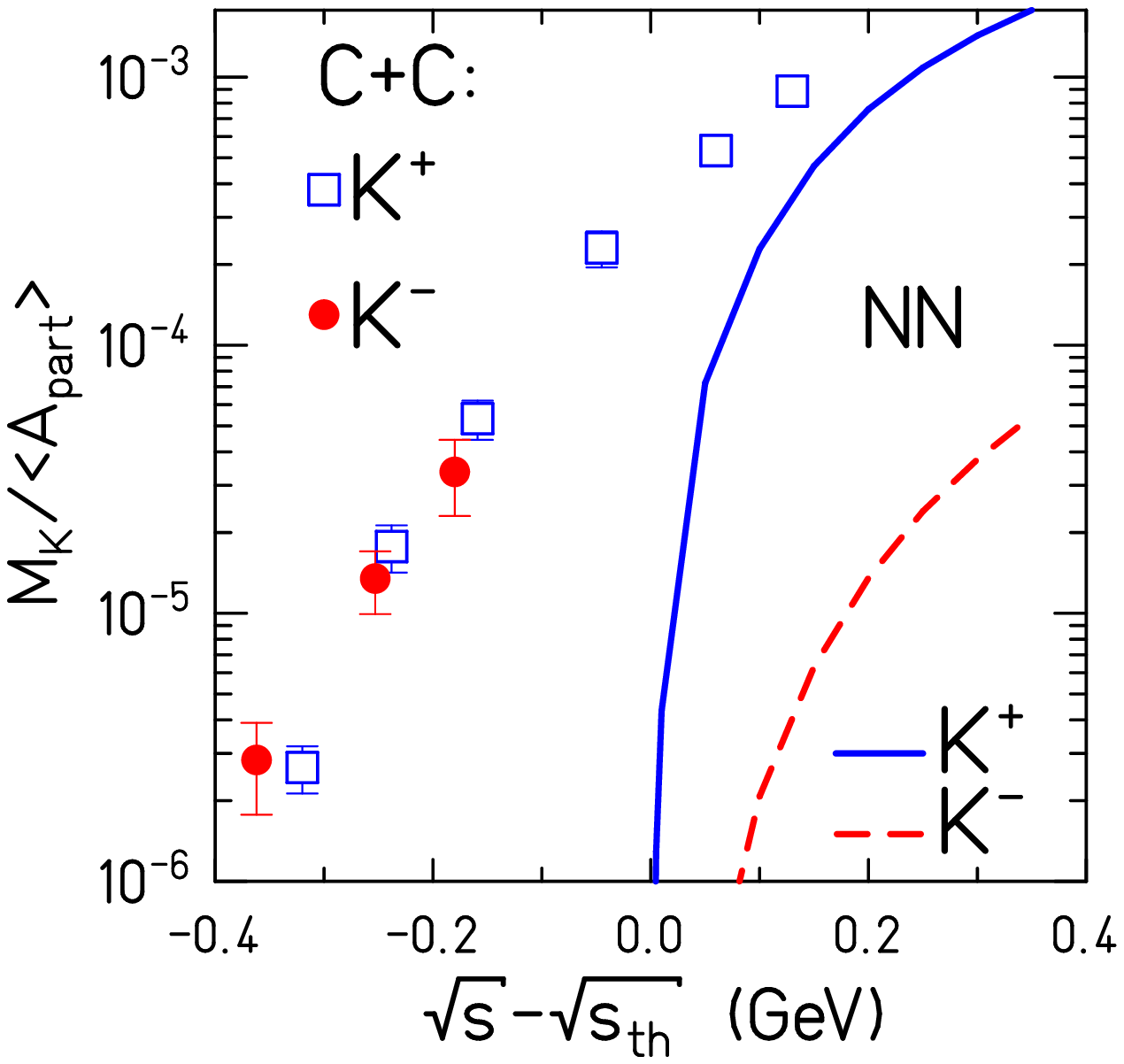,width=16.cm}}
\caption{}
\end{figure}

\begin{figure}[h]
\hspace{.0cm}\mbox{\epsfig{file=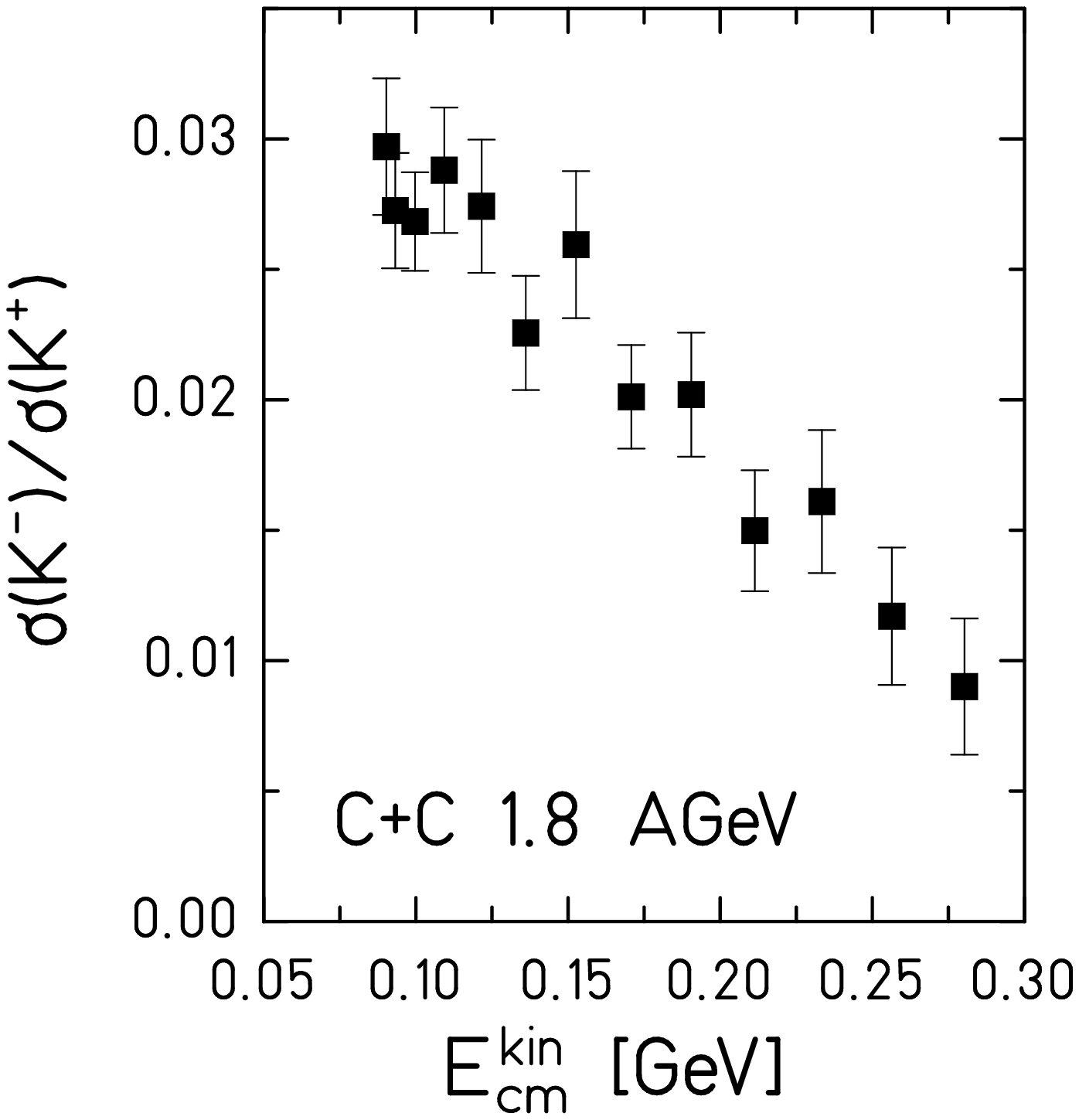,width=13.cm}}
\caption{}
\end{figure}

\end{document}